\documentclass[12pt]{article}
\usepackage{graphicx}
\usepackage{bm}
\newcommand\dpar[2]{\frac{\partial#1}{\partial#2}}
\def\be{\begin{equation}}
\def\ee{\end{equation}}
\def\ba{\begin{eqnarray}}
\def\ea{\end{eqnarray}}
\def\half{\frac{1}{2}}
\def\qmeta{q_m}
\def\Jmeta{J_m}
\def\bpsi{\bar{\psi}}
\def\det{\mbox{det}}
\def\sf{\sqrt{f}}
\def\D{\Delta}
\def\Drho{\Delta_{,\rho}}
\def\bma{\bm{\alpha}}
\def\cR{{\cal R}}
\def\cF{{\cal F}}
\def\barI{\bar{I}}
\def\barJ{\bar{J}}
\begin{document}
\begin{titlepage}
\vspace*{\bigskipamount}
\begin{center}
{\Large \bf Winding branes and persistent currents}
\end{center}
\vspace{\bigskipamount}
\begin{center}
Sergei Khlebnikov\footnote{\tt skhleb@physics.purdue.edu}
\end{center}
\begin{center}
\textit{Department of Physics, Purdue University, West Lafayette, IN 47907} 
\end{center}
\vspace{\bigskipamount}
\begin{center}
{\bf Abstract}
\end{center}
A D5 brane winding around a stack of D3 branes can be used as a model of persistent
current in a thin superconducting ring, with the number $N$ 
of D3s corresponding to the number of transverse channels in 
the ring. We consider, in the large $N$ limit, existence and properties 
of a gapped superconducting state with a uniform winding number density $q$. 
We find that there is a gapped classical solution for any $q$, no matter 
how large, but when $q$ is larger than a certain $\qmeta$ 
the state is unstable with respect to decay by phase slips. 
We argue that this decay
produces strings via a version of the Hanany-Witten effect (in a non-transverse, 
non-supersymmetric arrangement of branes). This parallels the requirement of quasiparticle 
production in a clean (disorder-free) wire in field theory of superconductivity.
\end{titlepage}
\setcounter{page}{2}
\tableofcontents
\section{Introduction}
In this paper, we consider solutions of type IIB supergravity that describe one
brane winding around another. The specific example we choose corresponds to 
the configuration in string theory that consists of a stack
of $N$ D3 branes and a single D5 brane, in the following
arrangement:
\be
\begin{array}{ccccccccccc}
& 0 & 1 & 2 & 3 & 4 & 5 & 6 & 7 & 8 & 9 \\
N \mbox{ D3s:} & \times & \times & \times & \times & & & & & & \\
\mbox{D5:} & \times & \times &  & & \times & \times & \times & \times &  & 
\end{array} 
\label{arr}
\ee
where crosses denote the coordinates wrapped by the branes. 
The only spatial direction shared
by all the branes is $x \equiv x^1$, which we assume to be a circle, possibly of
a large radius. We will be interested in the case when, as one goes around the circle,
the D5 winds around the D3s, the total of $W$ times.

Our interest in this configuration is that we believe it to be suitable for description
of a current-carrying state in a thin superconducting ring, with $W$ corresponding to the
winding number of the order parameter.
There are questions concerning thin superconductors, such as, for instance,
conditions of quantum stability of the supercurrent, that are
not easily answered by the conventional mean-field theory, and we would like to see
if supergravity can help us to answer these questions.

At small values of the 't Hooft coupling $\lambda = g_s N$ (where $g_s$ is the closed
string coupling), one may start by considering the
trivial embedding---with the D3s at $x^4=\ldots=x^9=0$, and the D5 at 
$x^2=x^3=x^8=x^9 =0$---and developing a field theory around it. This corresponds to 
replacing strings connecting the D5 to the D3s with their ground states and results in
a non-supersymmetric gauge theory of
$N$ species of left- and right-moving massless defect fermions. These correspond to
$N$ channels of conduction 
electrons in a wire.\footnote{
In a wire, different channels correspond to different transverse wavefunctions and 
different projections of spin. Thus, at large $N$, the material we are describing 
is one-dimensional only in its superconducting properties (e.g., the value of the order 
parameter), while the electrons are free to move in all three directions.}

The role of interactions, on the other hand, is most readily understood in the large
$N$ limit, when $g_s$ is small but $\lambda$ is large. In this case, it is
possible to replace the D3 branes by the classical black brane carrying $N$ units
of the D-brane charge \cite{Polchinski:1995mt}, i.e., by one of the solutions found
in \cite{Horowitz:1991cd}, while the D5 can be considered as a probe.
The near-horizon region of the D3 geometry has a field-theoretical interpretation 
even at strong coupling: it is dual to an $SU(N)$ superconformal gauge theory 
\cite{Maldacena:1997re,Gubser:1998bc,Witten:1998qj}. In our case, however, the D5 will
not stay in that region: once displaced from the trivial embedding, it is pushed out
to large absolute values of
\be
x^8 + i x^9 = \D e^{i\phi} \, .
\label{polar}
\ee
The correspondence of \cite{Maldacena:1997re,Gubser:1998bc,Witten:1998qj}
is not applicable at large $\D$, so supergravity is the only available
microscopic description of this strongly coupled system. It will be our starting point 
in what follows.

We interpret the instability of the trivial embedding as formation of a superconducting 
state \cite{Khlebnikov:2012ny}. The latter is
represented by a nontrivial embedding, which we find numerically for various values 
of the winding of the phase $\phi$. A nonzero winding corresponds to a nonzero 
supercurrent in the wire.
All embedding solutions we consider here are uniform in the
$x$ direction and so describe a superconductor
in the clean limit (i.e., without disorder). In particular, the gradient of $\phi$ 
(the winding number density, in units of $1/2\pi$) is independent of $x$ and given by
\be
q = \frac{2\pi W}{L} \, ,
\label{q}
\ee
where $L$ is the length of the wire. In practice, disorder is often important, and
the first step towards including it in our description would be to consider 
$x$-dependent background geometries. We discuss this further in the concluding section.

We find that there is a nontrivial embedding
solution for any value of $q$, no matter how large. For each of these solutions,
the minimal distance from the D5 to the D3s---the quasiparticle gap in units
of the string tension---is nonzero.
This is in contrast to the case of fixed current (and zero winding) we have considered
in \cite{Khlebnikov:2012ny}. In that case,
the superconducting solution is gapless at any nonzero current and disappears at a
critical current. The distinction between the fixed-winding and fixed-current cases
has a parallel in the conventional
mean-field theory, as we discuss in Sec.~\ref{sec:bdg}.

The existence of a gapped classical solution for any $q$ may seem surprising: the
prevalent situation in superconductivity is when, at
a certain $q=q_L$, quasiparticles with momenta antiparallel to the flow become gapless,
and the supercurrent becomes classically unstable. This stability bound is
known as Landau's criterion; we recall its derivation in Sec.~\ref{sec:bdg}. 
Note,
however, that our superconductor is perfectly uniform and therefore momentum-conserving, 
while the Landau
process is not. So, it is possible that an instability of this type will not show up until
we break momentum conservation by going over to an $x$-dependent background.

Next, we consider stability of the supercurrent with respect to phase slips.
These are events (either thermal \cite{Little} or quantum \cite{Giordano}) that remove 
a unit of winding from the supercurrent. Since reducing supercurrent releases momentum 
from it, a sink of momentum is required.  
Unlike in the Landau process, however, that sink may be a part of the system itself, 
so that no momentum needs to be supplied from the outside (e.g., by disorder). 
The existing microscopic theory 
of this effect \cite{quasip} is based on the Bogoliubov-de Gennes (BdG) equations, which, 
when interpreted liberally enough (as reviewed in Sec.~\ref{sec:bdg}), 
allow one to go one step beyond the conventional 
mean field, to include fluctuations of the order parameter. The result
is that, in a uniform system at zero temperature, 
a phase slip {\em must} produce a certain number of fermionic quasiparticles (with total
momentum parallel to the flow), and the
associated energy cost may block the process altogether. 

The liberal interpretation of the BdG equations, referred to above, treats the order 
parameter and the quasiparticles as separate quantum fields. One may be concerned that
this involves double counting, as both are ultimately made of the same electrons.
One may counter this by arguing that, as in any two-fluid model, the two fields simply
represent two different transport mechanisms (in our case, of the electric charge), and the
separation of the electrons into superconducting and normal should not be construed too
literally. Nevertheless, it is of interest to develop an approach in which the
superconducting and normal components are represented by different {\em types} of 
excitations---e.g., an elementary field vs. a soliton or a string. Dual gravity provides 
precisely this type of approach.

An important role in our calculation
is played by the Wess-Zumino (WZ) term in the D5 
action. We find that for any nonzero winding $W$ it causes the D5 to have an amount, 
equal to $N W$, 
of the charge conjugate to the worldvolume gauge field. Since a QPS changes $W$, it changes
the amount of charge and, by charge conservation, the difference has to be picked up by
excitations. We interpret this transfer of charge as a version
of the Hanany-Witten (HW) effect
\cite{Hanany:1996ie}: when a D5 brane crosses a D3, a fundamental string is produced. 
This version of the effect is similar but not identical to the one well known in 
the literature, in which the D3 and D5 have no common spatial directions and preserve some
supersymmetry \cite{Bachas:1997ui,Danielsson:1997wq,Bergman:1997gf}.
We find it remarkable that the present version completely parallels
the requirement of quasiparticle production obtained in field theory of superconductivity 
on the basis of the BdG equations.

While a phase slip reduces the energy of the supercurrent, to see if it is in fact 
energetically allowed, we must take into account the energy of the produced
quasiparticles. We find that single phase slips are
energetically forbidden until the winding number density reaches a certain
$\qmeta$. Thus, although a solution with
nonzero supercurrent exists for any $q$, at $q > \qmeta$ it is unstable with respect
to decay by phase slips. We discuss
some consequences of this in the concluding section.

\section{A review of the BdG formalism} \label{sec:bdg}

The BdG formalism is not used for any calculations in this paper. We review it here
as a reference point for later comparison with the results obtained using gravity
and also to illustrate the distinction between the fixed-winding
and fixed-current cases.

Consider the following Lagrangian, describing $N$ species of (1+1) dimensional fermions,
$\psi_n$, which interact with an order parameter field $\Phi$: 
\be
L_F = \sum_n \left( i \bpsi_n \gamma^\alpha \partial_\alpha \psi_n
- \half \Phi \bpsi_n \psi^c_n - \half \Phi^* \bpsi^c_n \psi_n
+ \mu \bpsi_n \gamma^0 \psi_n \right) \, .
\label{LF}
\ee
Here $\alpha=0,1$ and $n=1,\ldots, N$. 
The superscript $c$ denotes charge conjugation, and $\mu$ is a chemical potential.
Each $\psi_n$ is a two-component Dirac spinor.
We use the representation of the $\gamma$ 
matrices with  $\gamma^0 = \sigma_1$, $\gamma^1 = -i \sigma_2$, where $\sigma$s are 
the Pauli matrices. Thus, in the normal state ($\Phi = 0$), the upper
components of $\psi_n$ describe right-, and the lower left-, moving fermions.

The Dirac fermions $\psi_n$ can be related to the electron operators $a_{R,L}$ of 
a superconductor by taking
$N$ even and setting, for each $\nu = 1,\ldots,N/2$,
\be
\psi_{2\nu-1} = \left( \begin{array}{c} a_{R\nu\uparrow} \\ a_{L\nu\downarrow} \end{array} \right) ,
\hspace{3em}
\psi_{2\nu} = \left( \begin{array}{c} a_{R\nu\downarrow} \\ - a_{L\nu\uparrow} \end{array} \right) ,
\label{pack}
\ee
where an arrow denotes the projection of spin. The minus sign in the second instance 
in (\ref{pack}) causes $\Phi$ to
couple to a spin singlet, the case for most conventional superconductors. The index $\nu$ 
labels the {\em transverse channel}. It reflects the fact that, while the fields
$\psi_n$ are functions of time and $x$ only, the electrons in a superconductor move in
all three spatial directions and can be classified according to their transverse
wavefunctions. 

The BdG equations are obtained from (\ref{LF}) by variation with respect to $\psi_n$.
Our interpretation of these equations will be broader than the conventional 
one: in most applications $\Phi$ is considered as a classical background, but we will
allow quantum changes in $\Phi$. This is necessary if we wish to use (\ref{LF})
to describe quantum phase slips.

In mean-field theory, one can describe a steady supercurrent as a state with a uniformly
wound order parameter:
\be
\Phi(x) = \Phi_0 e^{i q x} \, ,
\label{wind}
\ee
where $\Phi_0 \neq 0$ is a constant. A quantum phase slip
is described by a configuration (instanton) that interpolates in the Euclidean time
between the state (\ref{wind}) and a similar state with a smaller $q$, corresponding to 
one fewer unit of winding. Such an instanton has one zero mode for each fermionic species
\cite{Jackiw:1981ee,Weinberg:1981eu}. An adaptation of the
argument of \cite{'tHooft:1976up} then shows that $N$ fermions
must be produced in the phase slip process \cite{quasip}. 

The quasiparticle spectrum near the ground state (\ref{wind}) can be conveniently obtained
by redefining the Fermi fields as follows:
\be
\psi_n (x.t) \to e^{\frac{i}{2} q x} \psi_n (x,t) \, .
\label{subst}
\ee
This unwinds $\Phi$ into a constant, $\Phi(x) \to \Phi_0$, but also modifies the Lagrangian:
\be
L_F \to L_F - \frac{q}{2} \sum_n \psi_n^\dagger \sigma_3 \psi_n \, .
\label{chem}
\ee
The additional term in (\ref{chem}) gives 
different chemical potentials to the right- and left-moving fermions. 
The corresponding Hamiltonian can be diagonalized by a Bogoliubov transformation.
The resulting excitation energies are
\be
\epsilon_{\pm}(k) =   [(k-\mu)^2 + |\Phi_0|^2]^{1/2} \pm \frac{q}{2} \, .
\label{disp}
\ee
The upper branch, $\epsilon_+(k)$, describes excitations with momentum $k$, and
the lower branch, $\epsilon_-(k)$, those with momentum $-k$. Both reach
minima at $k=\mu$, which can therefore be identified with the Fermi momentum: 
$\mu = k_F$. The lower branch touches zero when $q$ becomes equal to
\be
q_L = 2 |\Phi_0| \, ,
\label{qL}
\ee
that is twice the quasiparticle gap. Once $q$ exceeds $q_L$, it becomes energetically 
favorable to produce excitations with momenta near $-k_F$, and the ground state becomes
unstable. The stability condition $q<q_L$ is known as Landau's criterion.
As we have noted in the introduction (and discuss further in the conclusion), we 
do not recover this condition in the gravity-based calculation, presumably because
of the perfectly momentum-conserving nature of our system.

We can now define two limiting types of current-carrying states. 
In one type, quasiparticles are
absent but the order parameter is wound as in (\ref{wind}); the current this
state carries is a {\em supercurrent}. 
We refer to this case as {\em fixed winding}.
In the other type of state, $\Phi(x) = \Phi_0$, but the upper branch of the excitation
spectrum (\ref{disp}) is filled with quasiparticles, up to a certain finite density of them.
The current these carry is a {\em normal} current. Because the quasiparticles 
form a Fermi surface,\footnote{Not to be confused with the original electron Fermi surface
at $k=\mu$.} 
this state is gapless.  We identify it as a mean-field counterpart of 
the gapless superconductor described via a gravity dual in \cite{Khlebnikov:2012ny}. 
We refer to this case as {\em fixed current}.

In general, one must allow for both supercurrent and normal current components to be
present.
As long as phase slips are neglected, the momenta of these components are
separate conserved quantities. Phase slips, however,
convert winding of the order parameter into momentum of quasiparticles and vice versa.
The equilibrium winding and density of quasiparticles are then determined by comparing the free
energies of different states with the same total momentum. In this paper, we will be 
interested in a condition under which phase slips become energetically favorable and
the decay of the supercurrent-only state begins to populate the normal component.

\section{Computation of the WZ term} \label{sec:wz}
Equations of motion for the D3 geometry can be derived from the action
\be
S_{10} = \frac{1}{2 \kappa^2} 
\int d^{10} x \sqrt{-g} \left( \cR - \frac{1}{4\times 5!}  G_{(5)}^2 \right) .
\label{S10}
\ee
Here $\cR$ is the Ricci scalar, and $G_{(5)}$ is the Ramond-Ramond 5-form field strength. 
We use the shorthand notation
\be
G_{(5)}^2 = G_{ABCDE} G^{ABCDE} \, .
\label{F2}
\ee
$G_{(5)}$ is self-dual, which implies $G_{(5)}^2 = 0$ but, as is common practice, we
impose self-duality in the equations of motion, rather than directly in the action.

In the classical limit, a stack of $N$ extremal D3 branes is described by a solution
\cite{Horowitz:1991cd} to the equations of motion following from (\ref{S10}), with 
\be
ds^2 = \frac{1}{\sf} (-dt^2 + (dx^i)^2 ) + \sf (dr^2 + r^2 d\Omega_5^2) \, ,
\label{ds2}
\ee
($i=1,2,3$) and 
\be
G_{(5)} = Q (\epsilon_{(5)} + *\epsilon_{(5)}) \, ,
\label{G5}
\ee
where $d\Omega_5^2$ is the metric on the unit 5-sphere, and $\epsilon_{(5)}$ is the
volume 5-form on it. 
The metric function in (\ref{ds2}) is
\be
f = 1 + \frac{R^4}{r^4}
\label{f}
\ee
and the relation between the various parameters is
\be
Q = 16\pi g_s (\alpha')^2 N = 4 R^4 \, ,
\label{Q}
\ee
where $g_s$ is the closed string coupling.
The D3s themselves are hidden behind the degenerate horizon of (\ref{ds2}) at $r=0$.

Note that we are using the full D3 geometry, rather than the near-horizon limit; the latter 
would correspond to neglecting unity in comparison with $R^4/r^4$ in (\ref{f}). The physical
significance of this has been discussed in the introduction. Technically,
we need
the full geometry because a probe D5 placed in it 
will be repelled to large values of the radius $r$, $r_{\min} \sim R$. 

The probe D5 is governed by the action \cite{Polchinski:1998rr}
\be
S_{D5} = - \tau_5 \int d^6 \xi \sqrt{-\det(P[g]_{ab} + F_{ab})} + \tau_5 \int A \wedge P[G_{(5)}] \, ,
\label{SD5}
\ee
where $\tau_5$ is the brane tension, $\xi^a$ with $a=0,\ldots,5$ are coordinates on the
brane, $P$ denotes pullbacks to the brane worldvolume, $A_a$ is the worldvolume
gauge field, and 
\be
F_{ab} = \partial_a A_b - \partial_b A_a 
\label{Fab}
\ee
is its field strength. The WZ term is the second term in (\ref{SD5}).

In our case, the D5 wraps only three of the five angles appearing in (\ref{ds2}),
so it is convenient to rewrite the metric as
\be
ds^2 = \frac{1}{\sf} \left( - dt^2 + (d x^i)^2 \right) +
\sf \left( d\rho^2 + \rho^2 d\Omega_3^2 + d \D^2 + \D^2 d \phi^2  \right) \, ,
\label{ds2rho}
\ee
where $\Delta$ and $\phi$ are polar coordinates in the $(x^8,x^9)$ plane. Then,
\be
r^2 = \rho^2 + \D^2 \, .
\label{rho2}
\ee
In our calculation, the instability with respect to the D5 slipping off the equator
of the $S^5$, i.e., developing a non-zero expectation value of $\D$, corresponds to 
a pairing instability towards a superconducting state.

With these notations, the coordinates on the brane are
\be
\xi^a = (t,x,\rho,\bma) \, ,
\label{xi}
\ee
where $x\equiv x^1$, and $\bma=(\alpha^1,\alpha^2,\alpha^3)$ are the three angles spanning 
the $S^3$ in (\ref{ds2rho}). 
If we restrict attention to
embeddings with $x^2=x^3=0$, the general form of the embedding is
$\D=\D(\xi^a)$, $\phi= \phi(\xi^a)$, $A_a=A_a(\xi^b)$.
It is, however, consistent to restrict the class of embeddings further, to
\ba
\D & = & \D(\rho) \, , \label{emb1} \\
\phi & = & q x \, , \label{emb2} \\
A_t & = & A_t(\rho) \, , \label{emb3}
\ea
where $q$ is a constant, and all $A_a$ with $a\neq t$ are equal to zero. 

For this class of embeddings, the WZ term in (\ref{SD5}) becomes
\be
S_{WZ} = \tau_5 \int dt dx d\rho d^3 \alpha A_t P[G_{(5)}]_{x\rho\bma} \, .
\label{SWZ}
\ee
Only the magnetic part of (\ref{G5}) contributes to (\ref{SWZ}). The requisite 
components of $G_{(5)}$ are 
\ba
G_{\D\phi\bma}(\rho,\bma,\D) 
& = & -  \D \rho^3 \partial_\rho f(\rho,\D) \sqrt{g_3(\bma)} \, , \\
G_{\rho\phi\bma}(\rho,\bma,\D)  
& = &  \D \rho^3 \partial_\D f(\rho, \D) \sqrt{g_3(\bma)} \, , \\
G_{\rho\D\bma}(\rho,\bma,\D)  & = & 0 \, ,
\ea
where $g_3$ is the determinant of the metric on $S^3$ and
\be
f(\rho, \D) = 1 + \frac{R^4}{(\rho^2 + \Delta^2)^2}
\label{frho}
\ee
[cf. (\ref{f}) and (\ref{rho2})]. The requisite component of the pullback is
\be
P[G_{(5)}]_{x\rho\bma} = - q \left( G_{\D\phi\bma} \Drho + G_{\rho\phi\bma} \right) \, ,
\label{pback}
\ee
where we use the shorthand $\Drho\equiv \partial_\rho \D$. Substituting this into 
(\ref{SWZ}) and integrating over $\bma$, we obtain
\be
S_{WZ} = 2\pi^2 q \tau_5 R^4 \int dt dx d\rho A_t(\rho) \frac{d}{d\rho} \Pi(\rho,\D(\rho)) \, ,
\label{SWZ2}
\ee
where
\be
\Pi(\rho,\D) = \frac{\rho^4}{(\rho^2 + \D^2)^2} \, .
\label{Pi}
\ee

The full D5 action (\ref{SD5}) for embeddings of the form (\ref{emb1})--(\ref{emb3})
is
\be
S_{D5} = - 2\pi^2 \tau_5 \int dt dx d\rho  \sqrt{Z}
(1 + \Drho^2 - F_{t\rho}^2)^{1/2} + S_{WZ} \, ,
\label{SD5full}
\ee
where 
\be
Z(\rho,\D) = \rho^6 f(\rho,\D) [1 + q^2 \D^2 f(\rho,\D)] \, ,
\label{Z}
\ee
with $f$ given by (\ref{frho}), and $\D$ by $\D(\rho)$.

\section{Numerical solutions} \label{sec:sol}

We call an embedding solution {\em superconducting} if it is nontrivial (i.e., 
$\D(\rho)$ is not identically zero) and, in addition, 
\be
\D(\rho\to \infty ) = 0 \, .
\label{bc_inf}
\ee
The latter condition means that that there is no superconductivity
in the ultraviolet.\footnote{In that respect, perhaps a better, if a bit cumbersome,
name for these embeddings would be {\em spontaneously} superconducting.}
If the D5 brane does not cross the horizon, i.e.,
\be
\D(\rho =0) \neq 0 \, ,
\label{gap}
\ee
the minimal distance from it to the D3s is finite and so then
is the quasiparticle gap. We refer to such solutions as gapped.
As we will see, there is a gapped superconducting solution for any value of $q$. 
This is in contrast to the same system at fixed current \cite{Khlebnikov:2012ny}, where the 
superconducting solution at any nonzero current is gapless.\footnote{The 
solution at zero current is gapped and coincides with the $q=0$ solution found here.}

The case $q=0$ is 2-dimensionally Lorentz invariant and so is subject
to Coleman's 
theorem \cite{Coleman:1973ci} on nonexistence of Goldstone bosons in 2 dimensions.
This implies that the long-range order (LRO) represented by the nontrivial
profile of $\D$ is destroyed by quantum fluctuations. One may suspect that
the same is true also for the Lorentz non-invariant case $q\neq 0$.\footnote{
In a gapless Lorentz non-invariant superconductor, the low-energy 
fluctuations of the order parameter can be damped by gapless quasiparticles. 
We assume that a finite gap prevents
this mechanism of stabilizing an LRO from operating in the present case.}
The strength of quantum fluctuations, however, is suppressed by the large $N$ (essentially, 
by the D5's tension in the gravity description or by the thickness of the wire).
As a result, as $N$ increases, significant deviations from classical behavior occur 
at progressively larger spatial scales. Thus, our classical solutions faithfully represent 
the physics except at these largest scales. We also recall that superconductivity
does not require an LRO; it only requires that the superconducting {\em density} does not
renormalize to zero at large distances.

The equation of motion obtained from (\ref{SD5full}) by variation with respect 
to $A_t$ integrates into
\be
\frac{\sqrt{Z} F_{t\rho}}{(1 + \Drho^2 - F_{t\rho}^2)^{1/2}} + q R^4 \Pi = - J_0 \, ,
\label{const}
\ee
where $\Pi$ is given by (\ref{Pi}), and $J_0$ is the integration constant. 
Solving this algebraically for $F_{t\rho}$, we obtain
\be
F_{t\rho} = - \frac{J}{\sqrt{C}} (1 + \Drho^2)^{1/2} \, ,
\label{F}
\ee
where
\ba
J(\rho, \D) & = & q R^4 \Pi(\rho, \D) + J_0 \, , \label{J} \\
C(\rho, \D) & = & Z(\rho, \D) + J^2(\rho, \D) \, . \label{C}
\ea
The condition (\ref{gap}) requires that $J_0 = 0$: 
unless that is so, $F_{t\rho}(\rho=0) \neq 0$, and $A_t$ cannot be smooth at $\rho=0$. 

Substituting (\ref{F}) into the equation obtained by variation of (\ref{SD5full}) with 
respect to $\D$, we obtain
\be
\frac{d}{d\rho} \frac{\sqrt{C} \Drho }{(1 + \Drho^2)^{1/2}}
= (1 + \Drho^2)^{1/2}  \dpar{\sqrt{C}}{\D} \, ,
\label{eqm}
\ee
where $C$ is given by (\ref{C}) (and $J$ by (\ref{J}), with $J_0 = 0$).
Regularity of $\D(\rho)$ at $\rho=0$ together with (\ref{gap}) leads to 
the boundary condition
\be
\Drho(\rho = 0) = 0 \, .
\label{bc_zero}
\ee
Thus, (\ref{eqm}) has to be solved with the boundary conditions (\ref{bc_inf}) 
and (\ref{bc_zero}).
We do that numerically by shooting from $\rho =0$ with $\D(\rho=0)$ as a shooting
parameter. 

\begin{figure}
\begin{center}
\includegraphics[width=4.25in]{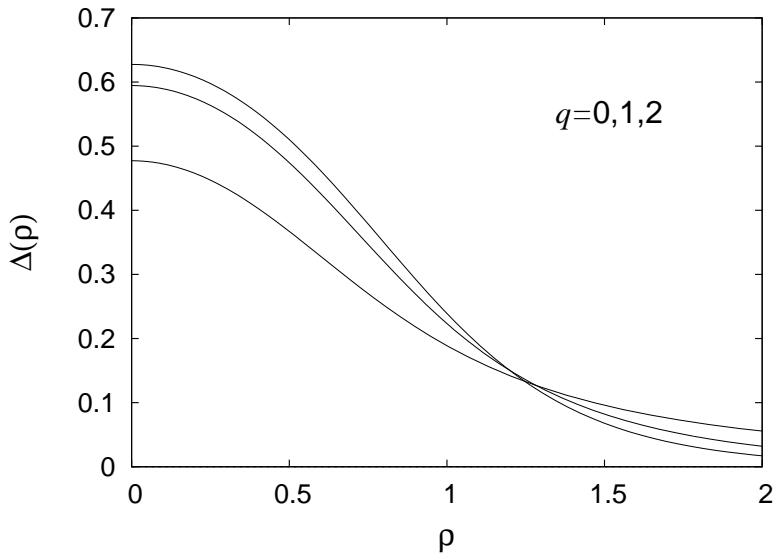}
\end{center}                                              
\caption{D5 profiles for three values of $q$. Both axes are in units of $R$, and $q$ is
in units of $1/R$. 
Larger $q$ correspond to larger values of
$\D(\rho=0)$.
}                                              
\label{fig:sol}                                                                       
\end{figure}

Some representative solutions are shown in Fig.~\ref{fig:sol}.
We find that such a solution exists for any $q$, no matter how large. 
The solution approaches a fixed shape as $q\to \infty$. 
The absence of a classical stability bound on $q$ (i.e., of the Landau criterion)
has been remarked upon in the introduction and is further discussed in the conclusion.
The minimal value of 
\be
r(\rho) = [\rho^2 + \D^2(\rho)]^{1/2} \, ,
\label{r}
\ee
which gives the quasiparticle gap in units of the string tension, is reached at
$\rho=0$. Thus, the gap is given by $\Delta(\rho=0)$. A somewhat counterintuitive result 
is that it grows with $q$ (and approaches a constant value at $q\to\infty$).

\section{Worldvolume flux and charge}
\label{sec:charge}

The counterpart of (\ref{const}) for the fixed current case \cite{Khlebnikov:2012ny} 
is obtained
by setting $q=0$ and retaining a nonzero $J_0$ (which is then interpreted as the fixed
current). Let us compare the fluxes of $F_{t\rho}$ at $\rho\to \infty$ for the two cases.
They are given by 
\ba
F_{t\rho}(\rho \to \infty) & = & - \frac{q R^4}{\rho^3} \hspace{3em} 
\mbox{(fixed $q$)} , \label{flux1} \\
F_{t\rho}(\rho \to \infty) & = & - \frac{J_0}{\rho^3} \hspace{3em} 
\mbox{(fixed current)} . \label{flux2}
\ea
These are obviously similar. Indeed,
as we will see shortly, $qR^4$ can be interpreted as the {\em average} current in the
fixed-$q$ case. Note, however, that, unlike (\ref{flux2}), which is sourced by charges 
behind the horizon, the flux (\ref{flux1}) is entirely due to charges on the brane.

According to (\ref{SWZ2}), the total charge,
coupling to the properly normalized potential $A_t / (2\pi \alpha')$, is
\be
4 \pi^3 \alpha' \tau_5 R^4 \int q dx = N W \, ,
\label{charge}
\ee
where $W= q L / 2\pi$ is the winding number, and we have used 
the expression (\ref{Q}) for $R$ and \cite{Polchinski:1998rr}
\be
\tau_5 = \frac{1}{(2\pi)^5 g_s (\alpha')^3} 
\label{T5}
\ee
for the brane tension. Note that (\ref{charge}) is an integer. In the context of the
original condensed-matter system, we identify it with the total momentum carried by
the current, in units of the Fermi momentum.

Consider a process (a phase slip) in which the winding number $W$ decreases by 
unity. According to (\ref{charge}), this releases $N$ units of charge from the brane. 
Since the charge is conserved, some other charged objects must appear. For a different
brane arrangement, where
a D3 and a D5 have no common spatial dimensions, it is known that 
a fundamental string stretching between the branes
is produced when they cross; this is the 
Hanany-Witten effect \cite{Hanany:1996ie} in one of its dual versions
\cite{Bachas:1997ui,Danielsson:1997wq,Bergman:1997gf}. Charge conservation leads us to
conclude that the same is true here. 

\begin{figure}
\begin{center}
\includegraphics[width=3.75in]{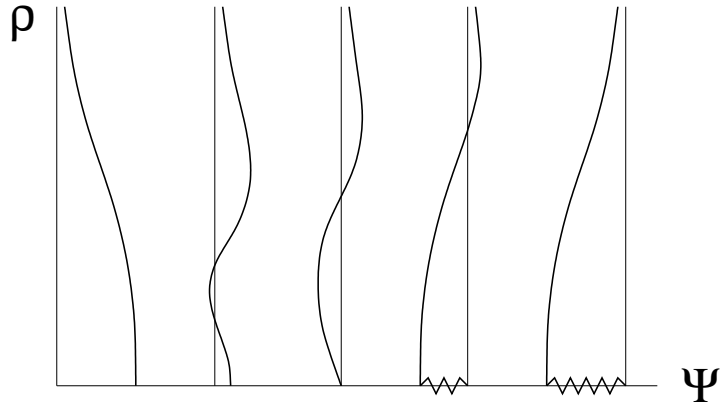}
\end{center}                                              
\caption{Schematic profiles of some $\Psi(\tau,x,\rho)$ that interpolates between
states connected by a phase slip. $\Psi = \Delta e^{i\phi}$ is the complex position 
of the D5 and is shown here at a fixed $x$ corresponding to the phase-slip center
(where it has been chosen real). Curves from left to right correspond to consecutive
values of the interpolation parameter $\tau$. They are shifted relative to one another,
so that all vertical lines correspond to $\Psi =0$. The D5 crossing the D3s 
(at $\Psi = \rho= 0$) corresponds to the profile in the middle. $N$ strings produced at
the crossing are shown by a wavy line.
}                                              
\label{fig:proc}                                                                       
\end{figure}

To visualize a phase slip, consider the complex position of the D5, 
$\Psi = \D e^{i\phi}$, as a function of $\rho$, $x$, and some interpolation parameter
$\tau$ (not necessarily the real time). Define the winding number
\be
W(\tau) = \frac{1}{2\pi} \int dx \partial_x \phi(\tau, x, \rho = 0) \, .
\label{W}
\ee
$W$ is a topological invariant (recall that we take the $x$ direction 
to be a circle): it will stay constant for small
fluctuations near a gapped solution. It can change, however, when the D5 passes 
through the D3s, i.e.,
$\D(\rho = 0) = 0$ at some values of $x$ and $\tau$.\footnote{In other words, 
the topological protection is incomplete, as fluctuations of the D5 can ``fill in''
the space between it and the D3s. Indeed, this is precisely 
the origin of phase slip processes.}
The process is shown schematically in Fig.~\ref{fig:proc}.
For generality, we consider the case when a zero of
$\D$ first appears at some nonzero $\rho$ and then propagates to $\rho = 0$;
alternatively, it may appear at $\rho = 0$ directly. When
the zero of $\D$ reaches $\rho=0$, the D5 crosses the D3s, and light open string
modes appear. We expect that this is the point where $N$ open strings required by 
charge conservation are produced. Since in our case
an open string in the D-brane description corresponds to a 
quasiparticle in the superconductor, we conclude
that each phase slip must produce $N$ such quasiparticles. This is precisely the result
we have obtained previously by instanton computations on the basis of 
the Bogoliubov-de Gennes equations \cite{quasip}.

\section{Computation of free energies} \label{sec:ene}

One consequence of quasiparticle production by a phase slip 
is that it affects the
energy balance: the energy gain from unwinding the supercurrent must 
be weighed against the energy cost of the produced quasiparticles. 
As a result, when production of quasiparticles is a requirement,
phase slips may be energetically forbidden (at zero 
temperature), and the supercurrent stable. In this section, we describe a numerical
computation of the relevant free energies.

Substituting (\ref{F}) into the expression (\ref{SD5full}) for the action and
integrating by parts in the WZ term, we obtain
\be
S_{D5} = 
2\pi^2 \tau_5 \int dt dx \left[ - \int d\rho \sqrt{C} (1+ \Drho^2)^{1/2} + q R^4 A_t(\infty)
\right] \, .
\label{SD5sol}
\ee
The integral over $\rho$ here is divergent at $\rho\to \infty$, but that can be fixed 
by subtracting some reference $q$-independent expression, e.g.,
\be
\sqrt{C} (1+ \Drho^2)^{1/2} \to \sqrt{C} (1+ \Drho^2)^{1/2}  - \sqrt{C_0} \, ,
\label{subtr}
\ee
where $C_0 = \rho^6(1 + R^4/\rho^4)$. 

The asymptotic value $A_t(\infty)$ appearing in (\ref{SD5sol})
can be interpreted as a chemical potential for the electric
current. This can be seen as follows. The average current 
carried by electrons near the Fermi surface is given by their momentum density 
in units of $k_F$. In the D-brane description, the momentum density is represented 
by the density of charge conjugate to $A_t$. Thus, changing the chemical potential 
for the current corresponds to a shift
\be
A_t(\rho) \to A_t(\rho) + \Lambda_0 \, ,
\label{shift}
\ee
where $\Lambda_0$ is a constant.
Allowing the D5 to have a nonzero total charge implies that the physical states of the
D5 are only invariant under gauge transformations with parameters vanishing at infinity,
in particular, $A_t(\rho) \to A_t(\rho) + \Lambda(\rho)$ with 
$\Lambda(\rho) \to 0$ at $\rho\to \infty$.
Then, $A_t(\infty)$ is a gauge-invariant quantity (a collective coordinate in the
terminology of \cite{Witten:1995im}), precisely the one modified 
by the shift (\ref{shift}).\footnote{Note that, unlike other identifications of
chemical potentials in D-brane systems (for example, the one in \cite{Apreda:2005yz}), 
the present argument does not rely on the AdS/CFT dictionary of 
\cite{Maldacena:1997re,Gubser:1998bc,Witten:1998qj}. That dictionary cannot be used here 
as $\rho\to\infty$ corresponds to the asymptotically flat region of the geometry.}

Variation of (\ref{SD5sol}) with 
respect to $A_t(\infty)$ gives the average current $\barI = 2\pi^2 \tau_5 \barJ$, where
\be
\barJ = q R^4 \, .
\label{barJ}
\ee
Legendre transforming the action with respect to $A_t(\infty)$ and peeling off 
$(-\int dt)$ gives the free energy of the supercurrent state
\be
\cF_{\rm super}(q) = 2\pi^2 \tau_5 L \Theta(q) \, ,
\label{fren}
\ee
where
\be
\Theta(q) = \int_0^\infty  d\rho 
\left[ \sqrt{C} (1+ \Drho^2)^{1/2}  - \sqrt{C_0} \right] \, .
\label{Theta}
\ee

\begin{figure}
\begin{center}
\includegraphics[width=4.25in]{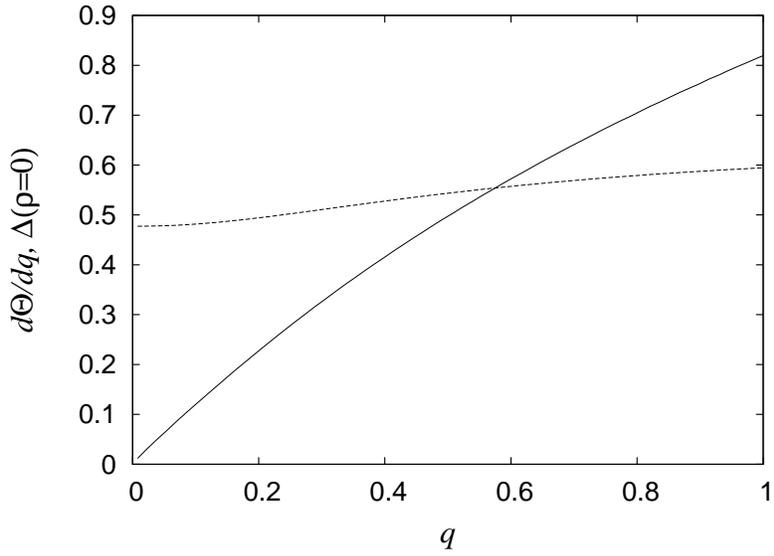}
\end{center}                                              
\caption{The amount of free energy (in units of $NR/2\pi \alpha'$; solid line) 
released by a single phase slip from the supercurrent in a long wire,  
as compared to the quasiparticle gap (in units of $R/2\pi\alpha'$; dashed line), 
both as functions of the winding number density $q$. Intersection of
the curves marks the value of $q$ (in units of $1/R$)
at which a phase slip becomes energetically allowed.
}                                              
\label{fig:comp}                                                                       
\end{figure}

Numerically computed function (\ref{Theta}) can be used for a partial analysis of 
stability of our solutions. 
In particular, consider classical stability of the solution
with respect to ``phase separation'': formation of a relatively large region along
$x$ where $\partial_x \phi$ is a constant different from $q$.
Classically, formation of such a region should start with a small 
($x$ and $t$-dependent) fluctuation of $\D$ and $\phi$ near the original uniform 
solution. As discussed in Sec.~\ref{sec:charge}, the total winding number 
can change only when $\Delta$ develops a zero at $\rho = 0$. For our solutions, 
$\Delta(\rho=0)$ is positive, and a small 
fluctuation will not change that. As a result, phase separation should begin
under the condition that the total winding number is unchanged: 
a decrease of $\partial_x \phi$
in a region of $x$ should be compensated by its increase elsewhere.
If the region is sufficiently large, its ends give only subleading 
contributions to the free energy, and the question of stability reduces to that of 
convexity of $\Theta(q)$.
The derivative $d\Theta / d q$ 
is plotted in Fig.~\ref{fig:comp}. It is monotonically increasing with $q$, which means
that the free energy is convex, and phase separation does not occur.

Even if (as the convexity argument suggests) the solution is stable with respect to small
fluctuations, it may still be able to decay by large ones---those that
allow $W$ to change (phase slips). 
We now consider various final states to which such a decay might lead.

The first group includes two gapless states: the normal state $\D \equiv 0$ and the gapless
superconductor found in \cite{Khlebnikov:2012ny}. A decay to either of these can be visualized 
as production of a finite density of strings that  pull
the D5 through the horizon, hiding behind it the worldvolume charge that has 
originally resided on the brane.
The momentum (charge) conservation implies that, to see if either decay 
is energetically allowed, we must compare the free energy (\ref{fren}) 
to that of the final state with the same value of the electric flux at infinity.
According to (\ref{flux1}) and (\ref{flux2}), this corresponds to setting $J_0$ for either
state equal to $\barJ$, eq.~(\ref{barJ}), of the supercurrent state. The free energies of
the gapless states are 
made finite in the same way as (\ref{Theta}); thus, for instance, the free energy of the
normal state is
\be
\cF_{\rm norm}(q) = 2\pi^2 \tau_5 L \int_0^\infty d\rho \left[ (C_0 + J_0^2)^{1/2} 
- \sqrt{C_0} \right] \, .
\label{fren_norm}
\ee
Numerically, we find that either of these free energies is 
always higher than that of the supercurrent state, so the decay is energetically 
forbidden.

Another type of final state is a gapped supercurrent state  with a smaller
value of $q$. Here we limit ourselves to states connected by a single phase slip,
$\Delta W = -1$, in a wire of macroscopic length $L$. In this case, a phase slip 
releases approximately $(2\pi/L) d\cF_{\rm super}/dq$ of free energy from the supercurrent.
It, however, requires production of $N$ quasiparticles. The minimal energy of a quasiparticle is
given by the quasiparticle gap, $(2\pi \alpha')^{-1}\Delta(\rho=0)$.
It is convenient to adopt the system of units where all lengths are measured in units of 
$R$, i.e., $R=1$. In these units, the coefficient in (\ref{fren}) can be written as
$2\pi^2  \tau_5 =N/ 4\pi^2 \alpha'$. So,
to see if a phase slip is energetically allowed,
we need to compare $d\Theta/dq$ to $\Delta(\rho=0)$. The comparison is shown
in Fig.~\ref{fig:comp}. We see that the decay becomes possible for $q$ larger 
than a certain $\qmeta$ (which is somewhat below 0.6, in units of $1/R$).

\section{Discussion} \label{sec:disc}
Overall, we find that supergravity provides a remarkably detailed picture of
a clean (disorder-free) multichannel  one-dimensional superconductor. The picture
includes the requirement of quasiparticle production by phase slips, seen here
as a version of the Hanany-Witten effect (in a non-transverse, non-supersymmetric 
arrangement of branes). 
This complements the earlier derivation \cite{quasip} based on
the Bogoliubov-de Gennes equations.

Gapped classical solutions exist in the present 
case for any winding number density $q$, no matter 
how large. As we have noted in the introduction, this may be a consequence of
momentum conservation, which precludes supplying a nonzero momentum to the 
superconductor as a whole (the Landau process). 
It would be interesting to see if gapped solutions seize to exist beyond a certain 
maximal $q$ in a case when momentum conservation is broken. 
One such case occurs when the stack of D3s is interrupted by an NS5 brane, 
in an arrangement similar to those considered in \cite{Hanany:1996ie}. The numbers
of D3s on the two sides of the NS5 need not be equal, so placing a probe D5 in such a 
geometry will represent a wire with a varying number of transverse channels, i.e.,
a constriction. The Landau process in this scenario would correspond 
to formation, at a critical current, of a region near the NS5 where quasiparticles 
with momenta antiparallel to the flow are gapless.
It remains to see, of course, if that is indeed what happens.

Unlike the Landau process, a phase slip does not need to change the total momentum
of the system: it can transfer 
momentum between the supercurrent and quasiparticles. We have seen that 
beyond a certain $q = \qmeta$
the supercurrent-only
solution is unstable with respect to decay by phase slips.
If we try to set up a supercurrent
larger than $\Jmeta = \qmeta R^4$ in a ring, a part of that current 
will decay into quasiparticles 
and, once these form a Fermi surface, a gapless normal component will appear. The latter 
will exist alongside a superconducting component---as we have seen in Sec.~\ref{sec:ene},
in the present case (a perfectly uniform wire at zero temperature),
the normal-only state is never the most energetically favorable. 

In the presence of disorder, phase slips can occur without quasiparticle 
production \cite{quasip}. Nevertheless, our present results lead us to consider the 
possibility that, in that case too, a sufficiently large current (now maintained by 
an external battery) causes appearance of a gapless normal component in 
the superconducting state.
One may contemplate trying to detect such a component experimentally---for instance, 
by measuring the current-voltage curve of electrons tunneling 
into the wire off the tip of a scanning tunneling microscope.

This work was supported in part by the U.S. Department of Energy grant \protect{DE-SC0007884}.


\begin{thebibliography}{99}
\bibitem{Polchinski:1995mt} 
  J.~Polchinski,
  ``Dirichlet branes and Ramond-Ramond charges,''
  Phys.\ Rev.\ Lett.\  {\bf 75}, 4724 (1995)
  [hep-th/9510017].
\bibitem{Horowitz:1991cd} 
  G.~T.~Horowitz and A.~Strominger,
  ``Black strings and P-branes,''
  Nucl.\ Phys.\ B {\bf 360}, 197 (1991).
\bibitem{Maldacena:1997re} 
  J.~M.~Maldacena,
  ``The Large N limit of superconformal field theories and supergravity,''
  Adv.\ Theor.\ Math.\ Phys.\  {\bf 2}, 231 (1998)
  [hep-th/9711200].
\bibitem{Gubser:1998bc} 
  S.~S.~Gubser, I.~R.~Klebanov and A.~M.~Polyakov,
  ``Gauge theory correlators from noncritical string theory,''
  Phys.\ Lett.\ B {\bf 428}, 105 (1998)
  [hep-th/9802109].
\bibitem{Witten:1998qj} 
  E.~Witten,
  ``Anti-de Sitter space and holography,''
  Adv.\ Theor.\ Math.\ Phys.\  {\bf 2}, 253 (1998)
  [hep-th/9802150].
\bibitem{Khlebnikov:2012ny} 
  S.~Khlebnikov,
  ``Critical current of a superconducting wire via gauge/gravity duality,''
  arXiv:1201.5103 [hep-th].
\bibitem{Little}
  W. A. Little,
  ``Decay of persistent currents in small susperconductors,''
  Phys. Rev. {\bf 156}, 396 (1967).
\bibitem{Giordano}
  N. Giordano,
  ``Evidence for macroscopic quantum tunneling in one-dimensional superconductors,''
  Phys. Rev. Lett. {\bf 61}, 2137 (1988).
\bibitem{quasip} S. Khlebnikov, 
  ``Quasiparticle scattering by quantum phase slips in one-dimensional superfluids,''
  Phys. Rev. Lett. {\bf 93}, 090403 (2004) 
  [cond-mat/0311045].
\bibitem{Hanany:1996ie} 
  A.~Hanany and E.~Witten,
  ``Type IIB superstrings, BPS monopoles, and three-dimensional gauge dynamics,''
  Nucl.\ Phys.\ B {\bf 492}, 152 (1997)
  [hep-th/9611230].
\bibitem{Bachas:1997ui} 
  C.~P.~Bachas, M.~R.~Douglas and M.~B.~Green,
  ``Anomalous creation of branes,''
  JHEP {\bf 9707}, 002 (1997)
  [hep-th/9705074].
\bibitem{Danielsson:1997wq} 
  U.~Danielsson, G.~Ferretti and I.~R.~Klebanov,
  ``Creation of fundamental strings by crossing D-branes,''
  Phys.\ Rev.\ Lett.\  {\bf 79}, 1984 (1997)
  [hep-th/9705084].
\bibitem{Bergman:1997gf} 
  O.~Bergman, M.~R.~Gaberdiel and G.~Lifschytz,
  ``Branes, orientifolds and the creation of elementary strings,''
  Nucl.\ Phys.\ B {\bf 509}, 194 (1998)
  [hep-th/9705130].
\bibitem{Jackiw:1981ee} 
  R.~Jackiw and P.~Rossi,
  ``Zero modes of the vortex-fermion system,''
  Nucl.\ Phys.\ B {\bf 190}, 681 (1981).
\bibitem{Weinberg:1981eu} 
  E.~J.~Weinberg,
  ``Index calculations for the fermion-vortex system,''
  Phys.\ Rev.\ D {\bf 24}, 2669 (1981).
\bibitem{'tHooft:1976up}
  G.~'t Hooft,
  ``Symmetry breaking through Bell-Jackiw anomalies,''
  Phys.\ Rev.\ Lett.\  {\bf 37} (1976) 8.
\bibitem{Polchinski:1998rr} 
  J.~Polchinski,  
  ``String theory. Vol. 2: Superstring theory and beyond,''
  Cambridge, UK: Univ. Pr. (1998) 531 p.
\bibitem{Coleman:1973ci} 
  S.~R.~Coleman,
  ``There are no Goldstone bosons in two-dimensions,''
  Commun.\ Math.\ Phys.\  {\bf 31}, 259 (1973).
\bibitem{Witten:1995im} 
  E.~Witten,
  ``Bound states of strings and p-branes,''
  Nucl.\ Phys.\ B {\bf 460}, 335 (1996)
  [hep-th/9510135].
\bibitem{Apreda:2005yz} 
  R.~Apreda, J.~Erdmenger, N.~Evans and Z.~Guralnik,
  ``Strong coupling effective Higgs potential and a first order thermal phase transition 
  from AdS/CFT duality,''
  Phys.\ Rev.\ D {\bf 71}, 126002 (2005)
  [hep-th/0504151].
\end{thebibliography}
\end{document}